\documentclass[conference]{IEEEtran}
\IEEEoverridecommandlockouts
\usepackage{cite}
\usepackage{amsmath,amssymb,amsfonts}
\usepackage{algorithmic}
\usepackage{graphicx}
\usepackage{textcomp}
\usepackage{xcolor}
\usepackage{url}
\usepackage{booktabs}
\usepackage{multirow}
\usepackage{listings}
\usepackage[breaklinks]{hyperref}
\usepackage{textcomp}
\usepackage{stfloats}
\usepackage{url}
\usepackage{verbatim}
\usepackage{graphicx}
\usepackage{cite}
\usepackage{tcolorbox}
\usepackage{tikz}
\usepackage{multirow}
\usepackage{listings}
\usepackage{xcolor}
\usepackage{tabularx}
\usepackage{colortbl}
\usepackage{caption}
\usepackage{pifont}

\usepackage{booktabs}

\definecolor{codebackground}{RGB}{246,242,241}
\definecolor{gitred}{RGB}{248,214,211} 
\definecolor{gitgreen}{RGB}{212,233,205}
\definecolor{javacomment}{RGB}{0,153,76}

\def\BibTeX{{\rm B\kern-.05em{\sc i\kern-.025em b}\kern-.08em
    T\kern-.1667em\lower.7ex\hbox{E}\kern-.125emX}}

\begin{document}

\title{Large Language Models are Qualified Benchmark Builders: Rebuilding Pre-Training Datasets for Advancing Code Intelligence Tasks

\thanks{This research was supported by the the National Key Research and Development Program of China (Grant No.2023YFB4503802), the National Natural Science Foundation of China (Grant No.62172426,62302515) and the NUDT Research Project for Student(No.ZC525Z042402).\\
$^\star$Xinjun Mao and Shangwen Wang are the corresponding authors.
}
}
\author{
\IEEEauthorblockN{
        Kang Yang\IEEEauthorrefmark{1},
        Xinjun Mao\IEEEauthorrefmark{1},
        Shangwen Wang\IEEEauthorrefmark{1},
        Yanlin Wang\IEEEauthorrefmark{2},  
        Tanghaoran Zhang\IEEEauthorrefmark{1},\\
        Bo Lin\IEEEauthorrefmark{1},
        Yihao Qin\IEEEauthorrefmark{1},
        Zhang Zhang\IEEEauthorrefmark{1},
        Yao Lu\IEEEauthorrefmark{1},
        Kamal Al-Sabahi\IEEEauthorrefmark{3}
    }

\IEEEauthorblockA{\IEEEauthorrefmark{1} College of Computer, National University of Defense Technology, Changsha, China,\\
    \{yangkang, xjmao, wangshangwen13, zhangthr, linbo19, yihaoqin, zhangzhang14, luyao08 \}@nudt.edu.cn
    }

\IEEEauthorblockA{\IEEEauthorrefmark{2} Sun Yat-sen University, wangylin36@mail.sysu.edu.cn
    }

\IEEEauthorblockA{\IEEEauthorrefmark{3} College of Banking and Financial Studies, Oman, kamal@cbfs.edu.om
}
}

\maketitle

\begin{abstract} 
Pre-trained code models are essential for various code intelligence tasks. Yet, their effectiveness is heavily influenced by the quality of the pre-training dataset, particularly human-written reference comments, which usually serve as a bridge between the programming language and natural language.
One significant challenge is that such comments could become inconsistent with the corresponding code as the software evolves, leading to suboptimal model performance.
Large language models (LLMs) have demonstrated superior capabilities in generating high-quality code comments. 
This work investigates whether substituting original human-written comments with LLM-generated ones can improve pre-training datasets for more effective pre-trained code models. 
As existing reference-based metrics cannot evaluate the quality of human-written reference comments themselves, to enable direct comparison between LLM-generated and human reference
comments, we introduce two auxiliary tasks as novel reference-free metrics, including code-comment inconsistency detection and semantic code search. 
Experimental results show that LLM-generated comments exhibit superior semantic consistency with the code compared to human-written reference comments. 
Our manual evaluation also corroborates this conclusion, which indicates the potential of utilizing LLMs to enhance the quality of the pre-training dataset.
Based on this finding, we rebuilt the CodeSearchNet dataset with LLM-generated comments and re-pre-trained the CodeT5 model. 
Evaluations on multiple code intelligence tasks demonstrate that models pre-trained by LLM-enhanced data outperform their counterparts (pre-trained by original human reference comments data) on code summarization, code generation, and code translation tasks.
This research validates the feasibility of rebuilding the pre-training dataset by LLMs to advance code intelligence tasks. It advocates rethinking the reliance on human reference comments for code-related tasks.
\end{abstract}

\section{Introduction}
\label{sec:introduction}
In the realm of code intelligence, pre-trained code models have significantly enhanced a spectrum of tasks such as code summarization~\cite{song2019survey}, code generation~\cite{wang2023natural, singh2023codefusion, wang2023two}, code search~\cite{gu2018deep, sun2022importance, wang2024fusing}, clone detection~\cite{lei2022deep} and automated bug fixing~\cite{lin2024one, lin2022context}. Pre-trained source code models such as CodeBERT~\cite{feng2020codebert}, GraphCodeBERT~\cite{guo2020graphcodebert},
CodeT5~\cite{wang2021codet5} and UniXcoder~\cite{guo2022unixcoder} have achieved remarkable results on various software engineering tasks and even outperform large language models when fine-tuned with domain-specific data \cite{li2024understanding}. 
The comments of the corresponding code snippets serve as a crucial bridge between the programming language (PL) and natural language (NL), providing contextual understanding pivots that are vital for pre-training the above models.
As such, the effectiveness of the pre-trained code models relies heavily on the quality of their pre-training datasets~\cite{sun2022importance}, which depend heavily on human reference comments.

However, this reliance on NL comments introduces a fundamental challenge: as software evolves, these comments often become mismatched with the code~\cite{steidl2013quality, ying2005source, wen2019large}, leading to semantic inconsistencies between the code and comment.
Previous research has highlighted such inconsistencies. Shi et al.~\cite{shi2022we} revealed that 41.9\% of the code summarization dataset TLC~\cite{hu27summarizing} is noisy, with 22.8\% of the comments being inconsistent with the corresponding code snippets. Consequently, inconsistent comments deteriorate the quality of the PL-NL dataset, which can degrade the training efficacy and performance of code models. Sun et al.~\cite{sun2022importance} identified that more than one-third of the comments in CodeSearchNet (Java)~\cite{husain2019codesearchnet} did not describe core functionalities. The model trained with noisy data faces severe performance degradation~\cite{sun2022importance}.

Recently, LLMs~\cite{roziere2023code, achiam2023gpt, hurst2024gpt} have demonstrated superior generation capabilities in generating high-quality code comments~\cite{guo2023exploring, liang2023large, geng2024large}.
In light of this, we are motivated to utilize LLMs to address the limitations faced by pre-trained code models. Specifically, we aim to answer the following question: \emph{Can we rebuild the pre-training dataset by substituting the original human-written comments with LLM-generated ones for training more effective pre-trained code models?}

To that end, we first conduct a comprehensive evaluation to compare LLM-generated comments with human-written reference comments. To ensure the representativeness of the review, both open source (e.g., Code Llama~\cite{roziere2023code} released by Meta, DeepSeek-Coder~\cite{guo2024deepseek} developed by DeepSeek AI, and StarCoder2~\cite{lozhkov2024starcoder} built by BigCode in collaboration with NVIDIA) and closed source LLMs (e.g., Text-davinci-003/GPT-3.5/GPT-4 released by OpenAI~\cite{achiam2023gpt}) are considered in this work. 
A fundamental problem faced by our study is the lack of appropriate metrics.
Specifically, traditional evaluation metrics for code summarization are reference-based, measuring the similarity between predicted and human-written reference comments. However, the underlying assumption of the reference-based evaluation is that reference comments are gold standard superior to other baselines~\cite{gao2023umse}.
A critical purpose of this work is to compare the quality of LLM-generated comments with human-written reference comments. Thus, we cannot directly apply off-the-shelf reference-based metrics to assess the quality of reference comments, which makes reference-free evaluation essential for our comparative comparison.

To address this challenge, we introduce two auxiliary tasks, code-comment inconsistency detection and semantic code search, to offer a more refined reference-free assessment of code comment quality.
The inconsistency detection task aims to identify inconsistency between comment and code, and the semantic code search task assesses the ability to retrieve the correct code snippet using its comment as a query. 
Our intuition is that a higher-quality comment would exhibit better semantic consistency with the corresponding code so that (1) it is less likely to be detected as inconsistent by a well-trained classifier and (2) it is expected to facilitate accurate retrieval of the associated code from a database when used as a search query. 
The reference-free evaluation results show that \textit{\textbf{comments generated by LLMs preserve better semantic consistency with the code than human reference comments}}. 
In addition, we conduct a human evaluation to rate the LLMs-generated comments generated and human-written ones. The human evaluation results validate the effectiveness of two proposed reference-free metrics and confirm that comments generated by LLMs preserve better semantic consistency with code than the human-written reference comments. This insight forms the basis for reconstructing the training dataset. 

Due to the data quality issues~\cite{sun2022importance} in the widely used CodeSearchNet dataset, we rebuild it by substituting the human-written reference comments in
CodeSearchNet~\cite{husain2019codesearchnet} with LLM-generated ones. We further extend our research by re-pre-training the widely used model CodeT5~\cite{wang2021codet5} with this rebuilt CodeSearchNet dataset and evaluating its performance across five downstream code intelligence tasks. 
The model trained with the LLM-rebuilt dataset exhibits superior performance in code intelligence tasks like code summarization, code generation, and code translation compared to the counterpart model trained with original datasets containing human reference comments. These findings affirm the potential of LLMs in improving the quality of training data for code intelligence tasks and underscore the need to reevaluate the longstanding reliance on human reference comments. 

This research makes the following contributions: 
(1) Innovative Evaluation Metrics: To our knowledge, we are the first to leverage code comment inconsistency detection and semantic code search as reference-free evaluation metrics for assessing code comment quality. 
(2) Empirical Validation of LLMs: Empirical results demonstrate that LLM-generated comments preserve better semantic consistency with code than human-written references, validating the quality and reliability of LLM-generated comments in software engineering.
(3) Exploration of LLMs in Dataset Construction: We demonstrate the efficacy of LLMs for creating high-quality code comment datasets, successfully generating over 2M comments using GPT-3.5-turbo to rebuild the CodeSearchNet dataset. This approach validates LLMs' potential for enhancing code intelligence training data.

\section{Preliminaries}
\label{sec:preliminaries}
This section relates to the metrics measuring code comment quality (Section~\ref{sec2.1: accessing}), as assessing the quality of LLM-generated comments is essential in this work. Also, we explore the effectiveness of LLM as benchmark builders, so the NL-PL pre-trained models are discussed (Section~\ref{sec2.2: pre-trained}).

\subsection{Assessing the Quality of Code Comments}
\label{sec2.1: accessing}
Quantitatively evaluating the quality of generated summaries is challenging. Typically, this involves comparing the model-generated summary to a reference summary. The similarity between the generated and reference summaries is then calculated as an indicator of quality. These metrics, known as reference-based, are classified into two main categories: n-gram overlap and semantic similarity~\cite{zhang2019bertscore, haque2022semantic}.

The dominant evaluation methods are traditional n-gram overlap-based metrics like BLEU~\cite{papineni2002bleu}, ROUGE~\cite{lin2004rouge} and METEOR~\cite{banerjee2005meteor}, initially utilized in the machine translation community to measure the predicted translations’ similarity to reference translations. They compute whether the same n-gram appear in the same order in both predictions and references.

The evaluation methods based on semantic similarity assess similarity in embedding space, providing ``partial credit'' for word matches, as termed by Wieting et al.~\cite{wieting2019beyond}. The application of semantic similarity to code summary assessment is supported by growing evidence of the insufficiency of mere word overlap~\cite{stapleton2020human, mahmud2021code}. 
Mahmud et al. recommend BERTScore for its effectiveness in capturing semantic similarities~\cite{mahmud2021code}, while Haque et al. find that cosine similarity using Sentence-BERT and Universal Sentence Encoder representations closely align with human judgments~\cite{reimers2019sentence, cer2018universal}.

However, reference-based metrics rely on human-written comments, often mined from software repositories, which may not always be high quality as software evolves~\cite{linares2015developers, wen2019large}. Additionally, with LLMs capable of generating high-quality comments, reference-based metrics may struggle to capture nuances, particularly when assessing if generated comments are equal to or better than human references.

\subsection{NL-PL Pre-trained Models}
\label{sec2.2: pre-trained}
A standard NL-PL pre-trained model first involves training a large-scale model on extensive unlabeled datasets using self-supervised objectives, then fine-tuning it for specific downstream tasks like code understanding and generation using task-specific loss functions. This paradigm, initially introduced in NLP communities~\cite{feng2020codebert, guo2020graphcodebert, wang2021codet5}, has been proposed, featuring variations in architecture and pre-training tasks.

CuBERT~\cite{kanade2020learning} and CodeBERT~\cite{feng2020codebert} were among the first NL-PL models, with CodeBERT being the first large pre-trained model for multiple programming languages. Both use a multi-layer bidirectional Transformer architecture~\cite{vaswani2017attention}, similar to BERT~\cite{kenton2019bert} and RoBERTa~\cite{liu2019roberta}. CodeBERT is pre-trained on the bimodal CodeSearchNet dataset~\cite{husain2019codesearchnet} with two objectives: Masked Language Modeling (MLM)~\cite{clark2020electra}. GraphCodeBERT~\cite{guo2020graphcodebert}, which shares CodeBERT's architecture, adds structural code features by incorporating data flow graphs (DFG), replacing RTD with DFG-specific tasks such as predicting data flow edges and aligning nodes, leading to improved performance in several downstream tasks.

Encoder-only models require an additional decoder for generation tasks, limiting their ability to leverage pre-training for such tasks. Conversely, GPT-C~\cite{svyatkovskiy2020intellicode} and CodeGPT~\cite{lu2021codexglue} employ unidirectional language modelling, which works well for code generation but is suboptimal for understanding tasks. Recent work has explored encoder-decoder models for both understanding and generation tasks. PLBART~\cite{ahmad2021unified}, based on the BART architecture~\cite{lewis2020bart}, is pre-trained on both NL and PL with denoising objectives. CodeT5~\cite{wang2021codet5} modifies the T5 model~\cite{raffel2020exploring} to include token type information for identifiers, supporting multi-task learning in downstream applications. UniXcoder~\cite{guo2022unixcoder} further extends these ideas, using a decoder-only approach for auto-regressive tasks like code completion and incorporating multi-modal inputs such as code comments and abstract syntax trees (AST) to enhance representation.

Instead of optimizing model architectures or designing domain-specific pre-training tasks, our study emphasizes the impact of pre-training data quality. We investigate how improvements in pre-training dataset quality can enhance the performance of pre-trained code models on downstream tasks.
\section{Evaluation of LLM-generated comments}
\label{sec:evaluation}
In this section, we comprehensively compare the 
LLM-generated comments and human-written reference comments. 
\subsection{Objective and Research Questions}
Traditional code comment evaluation relies on reference-based metrics using human-written comments as the gold standard, limiting direct quality comparisons between generated and human reference comments~\cite{gao2023umse}. 
To address this limitation, we propose a paradigm shift towards reference-free evaluation using extrinsic auxiliary tasks: code-comment inconsistency detection and semantic code search. These tasks assess the semantic alignment between code and comments without requiring human references. In this section, our study addresses two key research questions:

\textbf{RQ1}: \textbf{How effective are the proposed reference-free metrics in assessing the quality of code comments?} 
We explore the effectiveness of our newly proposed evaluation metrics. It examines the alignment of reference-free metrics with reference-based metrics and human judges, laying the foundation for a comparative evaluation of LLM-generated and human reference comments.

\textbf{RQ2: How does the quality of comments generated by LLMs compare to human-written reference comments?}
We examine whether LLM-generated comments can match or exceed the quality of human-written references, which is central to determining the practicality of employing LLMs to enhance training datasets for code intelligence tasks.

\subsection{Reference-free Evaluation Metrics}
\subsubsection{Code Comment Inconsistency Detection}
\label{ssbsec:ccid}
Code comment inconsistency detection (CCID) determines whether a comment is semantically misaligned with the corresponding code snippet \cite{wen2019large}. Since CCID is of immense practical use to software developers who have a vested interest in keeping their code bases easily readable, navigable, and as bug-free as possible, prior works proposed kinds of approaches for detecting inconsistencies \cite{ panthaplackel2021deep, steiner2022code}. 
\begin{figure}[htbp]
\centerline{\includegraphics[scale=0.45]{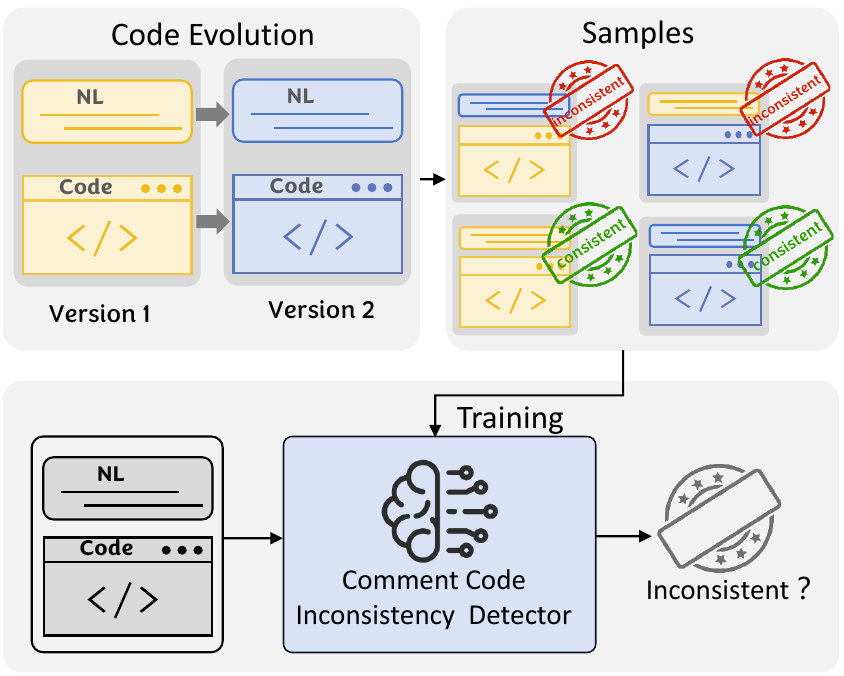}}
\caption{ Code comment inconsistency detection as a reference-free evaluation metric.}
\label{fig-inconsistency}
\end{figure}

We propose the CCID task as a quantitative measure of comment quality. This approach stems from the principle that high-quality comments should accurately reflect their corresponding code. Inconsistent comments can cause confusion, errors, or misinterpretations, reducing effectiveness. The CCID task identifies discrepancies between code and comments, serving as an objective metric for comment quality and providing insights into the effectiveness of software documentation. We evaluate comment quality by measuring inconsistency rates in test datasets.
The CCID task contains two distinct settings, post-hoc and just-in-time. In our work, we adopt the post-hoc setting because the comment/code pairs are available, and there are no code changes in the code summarization situation.

The data~\cite{ccidDataGoogleDrive} we used to train the CCID classifier is curated by Panthaplackel et al.~\cite{panthaplackel2021deep}, which includes 40,688 samples of @return, @param, and summary Javadoc comments paired with their corresponding Java code methods. They consider comment-code pairs from each version of consecutive commits: $(c_1, nl_1), (c_2, nl_2)$. 
We collect examples that code changes do exist between two versions in which $c_1$ $\neq$ $c_2$. 
As a result of code changes, the developer updated the comment because it would have otherwise become inconsistent. 
Therefore, if $nl_1$ $\neq$ $nl_2$, we take $nl_1$ comment to be inconsistent with $c_2$ code. As illustrated in Figure~\ref{fig-inconsistency}, $(c_1, nl_2)$ and $(c_2, nl_1)$ are consequently constitute the inconsistent examples. 
In contrast, if $nl_1$ = $nl_2$, the collected examples (i.e. $(c_1, nl_1)$ and $(c_2, nl_2)$) are labelled as consistent examples. The assumption is that the developer chose not to update the comment while modifying the code, as the comment was still consistent with the changes~\cite{panthaplackel2021deep}. 
Figure~\ref{fig-inconsistency} demonstrates the data constructing and training process of the CCID classifier. 

Specifically, we utilize the state-of-the-art approach in the post-hoc setting proposed by Steiner et al.~\cite{steiner2022code} as a classification model. The tokenized code $C$ and the $NL$ are concatenated into one sequence as input ([CLS] \textit{$C$ tokens} [SEP] \textit{$NL$ tokens} [SEP]), where [CLS] is the classification token and [SEP] is the separation token. The model outputs a binary label indicating whether the code-comment pair is inconsistent. Following the replication package~\cite{longformer4ccid}, the model achieved 87.21\% accuracy, 92.43\% precision, 80.69\% recall, and 86.16\% F1 score on the test data curated by Panthaplackel et al.\cite{panthaplackel2021deep}, and we utilize this model as a well-trained CCID classifier in this study. 

Suppose $f$ denotes the well-trained CCID classifier, and the input of $f$ is a code snippet $c$ and its corresponding comment $nl$. The output $f(c, nl) = 1$ indicates they are semantically inconsistent; otherwise, $f(c, nl) = 0$. Assume $<C, NL>$ is the code snippets, its paired model-generated/human-written comments in test datasets, and the total examples in test datasets is $N$. We define the inconsistency rate (\textit{IncRate}) by calculating the proportion of inconsistent examples. 
\begin{equation}
  IncRate=\frac{1}{N} \sum_{i=0}^{N}f(c_i,nl_i)
\end{equation}

We propose to use \textit{IncRate} as a reference-free metric to evaluate the quality of comments, and the lower \textit{IncRate} indicates better semantic consistency between code and comments.  

\subsubsection{Semantic Code Search}
\label{ssbsec:codesearch}
The semantic code search task is to retrieve a code snippet that matches a given query by effectively capturing the semantic similarity between the query and code. Semantic code search is a vital software development assistant significantly improving development efficiency and quality~\cite{gu2018deep}. 

Our motivation is rooted in the premise that the code search task could indicate the degree to which comments are aligned with their corresponding code. We assume that high-quality comments should not only elucidate code functionality but also enhance the discoverability of code snippets through semantic code search. This assumption aligns with the practical use case of developers who regularly rely on comments to navigate and understand large code bases. Additionally, code comments are an alternative to the practical queries in research communities~\cite{husain2019codesearchnet, sun2022importance}. Consequently, the effectiveness of comments in facilitating accurate and efficient code search results serves as a proxy for their quality, providing a concrete, measurable dimension to an otherwise subjective attribute of software documentation. Therefore, we argue comments that are more helpful for code search tasks are likely to be higher quality. Our study employs comments as queries within a code search task to assess the comment quality.
\begin{figure}[htbp]
\centerline{\includegraphics[scale=0.48]{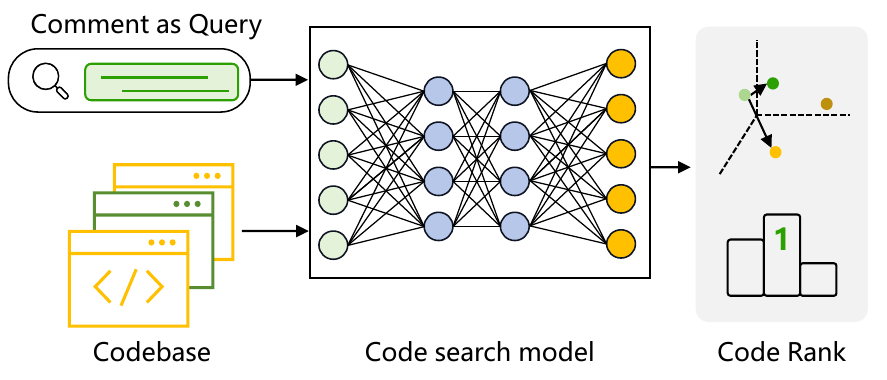}}
\caption{Code search as a reference-free evaluation metric.}
\label{fig-codesearch}
\end{figure}

To evaluate the performance of code search, we use the Mean of Reciprocal Rank (MRR) \cite{gu2018deep,lv2015codehow,ye2014learning}, which has been widely adopted in the evaluation of semantic code search. 
The MRR score quantifies the ranking of the target code snippet to the given comment query, and it only cares about where the most relevant result is ranked. 
We use CodeBERT~\cite{feng2020codebert} for code search task and follow the configuration reported in their artifacts~\cite{codesearchCodeBERT} to construct data and fine-tune the model. When computing MRR scores in the testing set, for each query-code pair, $|Q|-1$ code snippets from other pairs in the same batch play the role of distractor codes, where $|Q|$ is the batch size, and we set as $1000$ in this work. For a single batch, MRR is calculated as follows: 
\begin{equation}
  MRR = \frac{1}{\left | Q \right | } \sum_{i=1}^{\left | Q \right | } \frac{1}{Rank(\tilde{c}_i,nl_{i} )} 
\end{equation}
\noindent where $\tilde{c}_i$ is the ground-truth code snippet for its paired comment query $nl_{i}$, and $Rank(\tilde{c}_i,nl_{i} )$ is its corresponding rank in the retrieved results. MRR gives a score of the predicted result based on its rank. The average value of all batches is the final MRR score. 

We propose to use MRR as a reference-free metric to evaluate the quality of comments. A higher MRR indicates better semantic consistency between the code and the comments. 

\subsection{Experiment Design}
\subsubsection{Dataset}
\label{ssbsec:dataset}
We conduct experiments on a widely used Java benchmark dataset TLC \cite{hu27summarizing}, which has 66k code-comment pairs collected from more than 9K open-source Java projects created from 2015 to 2016 with at least 20 stars. They first extracted Java methods and Javadocs and treated the first sentence of the Javadoc as the ground-truth comment of the corresponding code. We use the TLC dataset open-sourced by the previous work \cite{shi2022we}. The training/validation/test set contains 53,528/7,555/4,985 samples, respectively.
\subsubsection{Baselines}
\label{ssbsec:baselines}
Our research aims to investigate the effectiveness of two proposed reference-free metrics, comprehensively analyze comment quality, and compare LLMs' performance with human-written comments. Three deep learning (DL) based code summarization models are introduced as baselines to provide a meaningful context for evaluation. 
Six LLMs, encompassing both open-source and closed-source models, are examined in this work to enhance the representativeness of the LLMs. This comparison will facilitate a more nuanced understanding of the performance differences in DL model-generated, LLMs-generated, and human-written comments.


\textbf{NCS} \cite{ahmad2020transformer} is a transformer-based model that utilizes relative distances in self-attention and includes a copy mechanism to handle rare tokens from source code. \textbf{SIT} \cite{wu2021code} integrates AST structure features into self-attention by combining the AST tree, data-flow graph, and control-flow graph into a multi-view graph matrix to refine token connections. \textbf{DOME} \cite{mu2023developer} leverages intent-guided selective attention to extract intent-relevant details, generating comments tailored to various intents. We use consistent data splits and default training parameters as in the original studies.


In this work, three types of models published by OpenAI are considered. \textbf{Text-davinci-003}. The text-davinci-003~\cite{textdavinci003} is a model trained using reinforcement learning with rewards from human comparisons. It performs well in consistent instruction following and longer output when completing any language task. \textbf{GPT-3.5 Turbo}. The capable and most cost-effective model (costs 1/10th of text-davinci-003) in the GPT-3.5 family can understand and generate natural language and code. The snapshot of GPT-3.5-turbo from Jan 25th, 2024, gpt-3.5-turbo-0125~\cite{gpt35t} is used in this work. \textbf{GPT-4 Turbo} can solve complex problems with greater accuracy than any previous models of OpenAI, thanks to its broader general knowledge and advanced reasoning capabilities. 
The snapshot gpt-4-0125-preview~\cite{gpt4} is utilized in this work.

The development process of LLMs varies in openness. Proprietary models, like OpenAI's GPT-4, are accessible via paid APIs, while their development specifics remain undisclosed. In contrast, open-source LLMs release model weights, allowing the community to run, inspect, and fine-tune these models. We consider three open-source LLMs:
\textbf{Code Llama} by Meta AI, based on Llama2~\cite{touvron2023llama}, provides state-of-the-art performance, extensive input context handling, and zero-shot instruction-following for programming tasks. This study uses the 70B-parameter Code-Llama model~\cite{code-llama}.
\textbf{DeepSeek Coder}~\cite{deepseekcoder} is a range of open-source code models trained on 2 trillion tokens from 87 programming languages, with a 16K context length for complex tasks. We employ the largest model, DeepSeekCoder-33B.
\textbf{StarCoder2}~\cite{starcoder2} by BigCode includes models with 3B, 7B, and 15B parameters trained on The Stack v2 dataset. All models and training resources are fully open-source.



To obtain summaries written by LLMs, we design a simple prompt with the format: ``\textit{You are an expert} [PL] \textit{programmer}. \textit{For the given} [PL] \textit{method}, \textit{please write a one-sentence description as comment}: [Code Snippet Content]"


\begin{table}[t!]
\centering
\caption{Statistics of unique words occurred in baseline-generated comments on TLC test set.}
\label{tab:tlc-stat}
\begin{tabular}{lcc}
\toprule
Methods      & Avg.words  & Unique words \\ \hline
NCS                     & 8.72  & 3,485 \\
DOME                    & 8.91  & 3,924 \\
SIT                     & 8.68  & 4,034 \\ \hline
Human References        & 11.76 & 5,854 \\ \hline
StarCoder2-15B	        & 13.80 & 5,420 \\
Text-davinci-003        & 13.49 & 6,392 \\ 
CodeLlama-70B           & 14.18 & 5,931 \\
DeepSeekCoder-33B       & 15.56 & 6,742 \\
GPT-3.5-Turbo	        & 15.22 & 6,721 \\
GPT-4-Turbo	            & 16.44 & 7,719 \\ 
\bottomrule
\end{tabular}
\end{table}
\noindent We collected summaries from LLMs using this prompt for 4,985 Java methods in the test set. As LLM tends to give much longer comments than human-written references, we set the parameter \textit{max\_tokens=30} to meet the same length level as human references for a fair comparison. Table~\ref{tab:tlc-stat} displays the word length statistics and unique words for LLMs-generated and reference comments. As for the sampling temperature parameters, we set the $top\_p=1$ and the $tempreture=1$.

\subsubsection{Evaluation Metrics}
\label{ssbsec:eval-metrics}
We must also report commonly used traditional metrics in research communities for comparison as we propose new evaluation metrics. 
We categorize the evaluation metrics into two main groups based on whether references are needed as a standard: (1) reference-free metrics (IncRate and MRR) and (2) reference-based metrics (BLEU, ROUGE, METEOR and USE).

\textbf{IncRate} is a reference-free metric for code summarization. As introduced in the former subsection~\ref{ssbsec:ccid}, IncRate evaluates the quality by measuring the percentage of inconsistent comment-code pair examples in the test set. 

\textbf{MRR} is used initially to evaluate information retrieval. As detailed in former subsection~\ref{ssbsec:codesearch}, we adopt it to measure the semantic relationship between the comment query and its paired code snippet. Higher MRR means a better quality of comment in semantic coherence.  


\begin{table*}[]
\centering
\caption{Evaluation of DL-based methods generated, 
LLMs-generated comments and human references comments.}
\label{tab:eval-cgpt}
\resizebox{\textwidth}{!}{
\begin{tabular}{l|cccc|cc|cccc}
\toprule
\multicolumn{1}{c|}{\multirow{2}{*}{Methods}} 
    & \multicolumn{4}{c|}{reference-based metrics} 
    & \multicolumn{2}{c|}{reference-free metrics} 
    & \multicolumn{4}{c}{human-evaluation  Avg. (Std.)}  \\
\multicolumn{1}{c|}{}                         
    & BLEU    & ROUGE  & METEOR  & USE     
    & InRate \(\downarrow\)            & MRR             
    & \multicolumn{1}{l}{Naturalness} 
    & \multicolumn{1}{l}{Consistency} 
    & \multicolumn{1}{l}{Usefulness}
    & \multicolumn{1}{l}{Average} \\ \hline
NCS                                           
    & 21.55   & 35.98  & 15.08   & 0.5198  
    & 31.84\% & 0.5614 
    & 3.31 (0.66)   & 3.19 (0.52)   & 3.36 (0.71)   & 3.29 \\
DOME                                          
    & 22.20   & 36.67  & 16.47   & 0.5460  
    & 24.99\% & 0.6296 
    & 3.27 (0.51)   & 3.23 (0.74)   & 3.47 (0.65)   & 3.32 \\
SIT                                           
    & \textbf{22.54}   & \textbf{37.87}  & 16.11   & 0.5634  
    & 23.65\% & 0.6404 
    & 3.35 (0.73)   & 3.28 (0.61)   & 3.44 (0.69)   & 3.36 \\ 
    \hline
Human References                              
    & -       & -      & -       & -       
    & 15.05\% & 0.8165 
    & 3.41 (0.82) & 3.35 (0.73) & 3.67 (0.91) & 3.48 \\ 
    \hline
StarCoder2-15B                                
    & 14.59   & 32.29  & 15.68  & 0.6297 
    & 3.95\%  & 0.8750 
    & 3.59 (0.78) & 3.44 (0.80) & 3.70 (0.86) &  3.58 \\  
Text-Davinci-003                             
    & 17.49   & 34.69  & 15.38   & 0.6234  
    & 3.65\%  & 0.8826 
    & 3.73 (0.85) & 3.57 (0.91) & 3.77 (0.81) & 3.69 \\ 
CodeLlama-70B                                 
    & 14.73   & 32.51  & 16.19  & 0.6353  
    & 2.91\%  & 0.9007 
    & 3.66 (0.83) & 3.70 (0.68) & 3.79 (0.92) & 3.72 \\
DeepSeekCoder-33B                             
    & 12.39   & 29.71  & 16.06  & 0.6242  
    & 2.31\%  & 0.9214 
    & 3.95 (0.75) & 3.79 (0.75) & 3.80 (0.83) & 3.85 \\ 
GPT-3.5-turbo                                 
    & 12.91   & 30.56  & 16.45   & \textbf{0.6505}  
    & 1.30\%  & 0.9562 
    & 3.98 (0.71) & 3.96 (0.88) & 3.95 (0.81) & 3.96 \\
GPT-4.0-turbo                                   
    & 11.24   & 28.61  & \textbf{16.63}   & 0.6434  
    & \textbf{0.52\%}  & \textbf{0.9679} 
     & \textbf{4.10} (0.67)& \textbf{4.02} (0.68) & \textbf{3.96} (0.95) & \textbf{4.03} \\ 
\bottomrule
\end{tabular}
}
\end{table*}

\textbf{BLEU} \cite{papineni2002bleu} is a textual similarity metric that calculates the precision of n-grams in a translated sentence compared to a reference sentence, with a weighted brevity penalty to punish short translations. We use the standard BLEU score, which provides a cumulative score of uni-, bi-, tri-, and quat-grams. 

\textbf{ROUGE} \cite{lin2004rouge} is a popular automatic evaluation metric that is recall-oriented. It computes the count of several overlapping units such as n-grams, word pairs, and sequences. ROUGE has several different variants, which we consider the ROUGE-L. 

\textbf{METEOR} \cite{banerjee2005meteor} is a metric based on the general concept of unigram matching, and it combines precision, recall, and a custom score determining the degree to which words are ordered correctly in the translation. 

\textbf{USE} \cite{haque2022semantic} is a metric that encodes the reference and the predicted summary to a fixed-length vector using a universal encoder and computes the similarity scores between two summaries. In this work, our experiments use the \textit{pre-trained universal-sentence-encoder-large} model \cite{cer-etal-2018-universal}. 

\subsection{Automatic Evaluation Results Analysis}
\subsubsection{reference-based metrics}
From the first three rows of Table~\ref{tab:eval-cgpt}, the performance of NCS, DOME, and SIT models appears closely matched, with no more than 2\% difference in BLEU, ROUGE, and METEOR scores. Such marginal disparities challenge the intuitive differentiation of model efficacy, and less than 2 points improvements of overlap-based metrics do not guarantee systematic improvements in summarization quality~\cite{roy2021reassessing}. 
When measured using reference-based evaluation metrics, the USE reveals a more significant discrepancy among the three DL-based models than BLEU, ROUGE, and METEOR. 
This phenomenon highlights the limitations of traditional n-gram overlap-based metrics in capturing subtle semantic differences. 

The traditional n-gram overlap-based metrics merely measure the literal proximity of predicted comments to reference comments. Yet, many words have close synonyms and certain words within a sentence carry more weight than others~\cite{haque2022semantic}. 
While comments generated by three DL-based models may seem similar on a superficial literal level, the USE metric, which assesses semantic similarity through embedding vectors, provides a nuanced ability to distinguish the quality differences between these DL models' generated comments.

\newtcolorbox{F1-box}{
colframe = gray!0!black,
boxrule = 0.6pt,              
top = 1pt,                  
bottom = 1pt,               
left = 3pt,                 
right = 3pt,               
boxsep = 1pt}
\begin{F1-box}
\textit{Finding 1: } Compared to n-gram overlap metrics, the USE metric demonstrates superior nuanced ability in capturing quality differences of DL baselines generated comments.
\end{F1-box}

\subsubsection{reference-free metrics}
Among three DL-based baselines, the SIT-generated comments exhibit the lowest IncRate, 23.65\%, DOME performs the second, 24.99\%, and NCS-generated comments get the highest inconsistency rate, 31.84\%, with their corresponding code snippets. Simultaneously, SIT-generated comments achieve the highest MRR scores, 0.6404, 
while DOME ranks second, 0.6296, and CSN performs the lowest MRR score, 0.5614. 
Our proposed reference-free metrics, IncRate and MRR scores, align well with the four traditional reference-based metrics when evaluating three DL-based models.

Furthermore, our proposed two reference-free metrics can reveal more substantial performance differences among NCS, DOME, and SIT than reference-based metrics. While n-gram-based metrics show less than a 2\% difference and USE metrics less than 5\%, the IncRate and MRR metrics demonstrate a wider gap of 8\%. It highlights the effectiveness of IncRate and MRR in detecting subtle but crucial semantic differences in comment quality across the DL-based models.

\newtcolorbox{F2-box}{
colframe = gray!0!black,
boxrule = 0.6pt,              
top = 1pt,                  
bottom = 1pt,               
left = 3pt,                 
right = 3pt,               
boxsep = 1pt}
\begin{F2-box}
\textit{Finding 2: } The reference-free metrics, IncRate and MRR, align well with reference-based metrics and demonstrate greater sensitivity in detecting performance differences in measuring DL baselines.
\end{F2-box}


\subsubsection{LLMs models vs DL models}
In comparing LLMs with three DL-based baseline models, all six LLMs fall behind in BLEU, ROUGE, and METEOR scores but surpass all three DL baselines in the USE metric. This discrepancy indicates that relying solely on reference-based metrics could yield unreliable or contradictory assessments. Notably, as shown in Table~\ref{tab:tlc-stat}, most LLMs (5/6) exhibit a richer vocabulary in their comments than human written reference comments. The DL-based models share the same vocabulary list with human references. It suggests that LLMs-generated comments preserve more semantic diversity in word/token selection, as they are usually pre-trained in a massive corpus. 
Therefore, the large vocabulary underscores the importance of semantic measurement over literal overlap for a more comprehensive evaluation of LLM's capabilities.

Based on the USE metric, all six LLM-generated comments show better likeness with human references than three DL-based baselines. According to our reference-free metrics, six LLMs achieve lower IncRate and higher MMR scores than all three DL-based baselines, which aligns well with the conclusion obtained by the USE metric. These results validate the effectiveness of IncRate and MRR in measuring the semantic likeness between code snippets and comments.

\newtcolorbox{F3-box}{
colframe = gray!0!black,
boxrule = 0.6pt,              
top = 1pt,                  
bottom = 1pt,               
left = 3pt,                 
right = 3pt,               
boxsep = 1pt}
\begin{F3-box}
\textit{Finding 3: } The comments generated by LLMs show better semantic likeness with human references than DL baselines.
\end{F3-box}

\subsubsection{LLMs vs human references} 

Reference-based evaluation metrics cannot evaluate the quality of the human reference comments themselves. Thus, they can not be used to compare the quality of comments on LLMs and human references. In contrast, our proposed metrics, IncRate and MRR, evaluate comment quality independently of references, which provide an objective basis for assessing comment quality across code summarization models, LLMs, and human developers.

As shown in Table~\ref{tab:eval-cgpt}, 15.05\% of human reference comments are detected as semantically inconsistent by the inconsistency detection classifier, suggesting that non-standard practices of co-evolving and maintaining comments with code often exist in software development and evolution. In comparison, the IncRate for six LLM-generated comments is markedly lower, no more than 4\%. GPT-4-Turbo achieves the lowest 0.52\% inconsistency rate. Additionally, in the code search task where comments are served as queries, all six LLM-generated comments surpass human references. GPT-3.5-Turbo and GPT-4-Turbo achieve MRR scores of 0.9562 and 0.9679, outperforming the human reference MRR score of 0.8165. 
\newtcolorbox{F4-box}{
colframe = gray!0!black,
boxrule = 0.6pt,              
top = 1pt,                  
bottom = 1pt,               
left = 3pt,                 
right = 3pt,               
boxsep = 1pt}
\begin{F4-box}
\textit{Finding 4: } LLM-generated comments preserve better semantic consistency than human reference comments.
\end{F4-box}



\subsection{Human Evaluation}
The proposed two reference-free metrics can measure the semantic consistency and semantic similarity between code snippets and comments without relying on human-written references, but it's not clear how they align with human judgment. To further validate the effectiveness of IncRate and MRR metrics, we perform a human evaluation to assess the quality of human reference and baseline-generated comments.

To evaluate the large number of instances across six LLMs and human references, we adopt a sampling method~\cite{singh2013elements} to ensure a reliable confidence interval. The minimum number of cases required is calculated using $MIN = n_0/(1+(n_0-1)/N)$, where $N$ is the total number of test cases, and $n_0 = (Z^2 \times 0.5)/e^2$, with $Z$ as the z-score for the desired confidence level and $e = 0.05$ at a $95\%$ confidence level.
Using this method, we selected 357 samples from a total of 4985 cases.
For each code snippet, its ten corresponding comments (three DL-model-generated, one human reference and six LLM-generated comments) are evaluated. We recruited 36 participants (24 Undergraduate, 8 Master's, and 6 PhD students) with over three years of programming experience in Software Engineering or Computer Science to manually assess the quality of these comments.


Following the previous study~\cite{su2023semantic, lin2023cct5, geng2024large}, participants rated each comment based on three aspects: 
(1) \textbf{Naturalness} reflects the fluency of comments from the grammar perspective.
(2) \textbf{Consistency} reflects the degree to which the semantics of comments match the code.
(3) \textbf{Usefulness} reflects the practical usage value for developers and how comments can help them. 
Scores range from 1 to 5 (1 for poor, 2 for marginal, 3 for acceptable, 4 for good, and 5 for excellent). For each code snippet, participants evaluated a reference comment, three comments generated by DL-based approaches, and six by LLMs. To ensure fairness, all ten comments were anonymized, and each participant completed the questionnaire independently. 
To avoid subjective bias, each comment was evaluated by two participants, and their average score was used. If their scores differed by two or more points, a third participant's evaluation was introduced to solve the conflict and determine the final score, as we consider a difference of more than 2 points to be an enormous disagreement.

Human evaluation results are presented in the third column of Table~\ref{tab:eval-cgpt}. The manual rating scores of human reference comments in Naturalness, Consistency and Usefulness are 3.41, 3.35 and 3.67, respectively, surpassing the three DL baseline models but scoring lower than all six LLM-generated comments in three aspects. The GPT-4-Turbo achieves the highest scores from participants, 4.10, 4.02 and 3.96 across these three aspects. In addition, the average score of humans rated in naturalness, consistency and usefulness is shown in the last sub-column. Evaluation results on two reference-free metrics aligned well with the overall quality level of manual evaluation. We thus answer the first two RQs (\textbf{RQ1\&RQ2}). 
\newtcolorbox{RQ1-box}{
colframe = gray!50!black,
top = 2pt,                  
bottom = 2pt,               
left = 3pt,                 
right = 3pt,               
boxsep = 2pt}
\begin{RQ1-box}
\textbf{Ans. to RQ1: effectiveness of reference-free metrics} \\
Our proposed reference-free metrics, IncRate and MRR, demonstrate a strong correlation with both four traditional reference-based metrics and human evaluations, indicating their effectiveness in assessing code comment quality.
\end{RQ1-box}

\newtcolorbox{RQ2-box}{
colframe = gray!50!black,
top = 2pt,                  
bottom = 2pt,               
left = 3pt,                 
right = 3pt,               
boxsep = 2pt}
\begin{RQ2-box}
\textbf{Ans. to RQ2: LLMs-generated vs. reference comments} 

The comments generated by LLMs exhibit higher quality than human reference comments, which preserve lower inconsistency and higher semantic relevance to the corresponding code. Human evaluations confirm that LLM-generated comments demonstrate superior naturalness, consistency and usefulness.
\end{RQ2-box}

\section{Distilling LLM for Code Intelligences}
\label{sec:distillation}
In pre-trained source code models, the quality of training data is crucial for model performance. PL-NL paired data helps bridge the semantic gap between programming and natural language, providing context and descriptive insight that facilitate code-intelligence tasks. Section~\ref{sec:evaluation} reveals that LLM-generated comments exhibit higher semantic relevance and lower inconsistency with the corresponding code snippets than human reference comments. 
These findings suggest that incorporating LLM-generated comments could enhance training data quality, motivating the reconstruction of pre-training datasets for code intelligence.
Thus, we ask the following research question:
\textbf{RQ3: How does the pre-training data rebuilt by LLM impact the performance of the downstream code intelligence tasks?} 

\subsection{Experiment Design}

To assess the impact of LLM-rebuilt data on code intelligence tasks, we conduct the following experiments: First, we reconstruct a popular dataset by replacing human-written comments with LLM-generated ones to improve semantic consistency between NL comments and PL code. We then use this rebuilt dataset to pre-train a widely used source code model, fine-tune it on downstream tasks, and evaluate its performance. Finally, we compare the results of models trained on the rebuilt versus the original dataset, quantitatively and qualitatively.
\begin{table}[]
\centering
\caption{Statistics of cgpt-CSN dataset.}
\label{tab:csn-stat}
\begin{tabular}{c|cc}
\toprule
PLs        & W/ ChatGPT-NL      & W/o NL  \\ \hline
Ruby       & 53,269    & 110,551  \\
JavaScript & 138,577   & 1,717,933 \\
Go         & 346,333   & 379,103  \\
Python     & 457,429   & 657,030  \\
Java       & 496,651   & 1,070,271 \\
PHP        & 578,072   & 398,058  \\ \hline
Total      & 2,070,331 & 4,332,946 \\ 
\bottomrule
\end{tabular}
\vspace{0.1cm}
\end{table}

\subsubsection{Dataset and Experimental Settings}

As the empirical results in Section~\ref{sec:evaluation}, the comments generated by GPT-4-turbo exhibit the best quality among six LLMs. The performance of GPT-4-turbo and GPT-3.5-turbo shows a minor difference in metrics, while the expenses of GPT-4-turbo are ten times more expensive than GPT-3.5-turbo. Due to consideration of OpenAI API costs, we rebuild the CodeSearchNet~\cite{husain2019codesearchnet} by replacing the human reference comments with GPT-3.5-turbo generated ones, noted as cgpt-CSN. The prompt we used is shown in Section~\ref{ssbsec:baselines}. We did not incorporate the C/CSharp dataset added in the original CodeT5~\cite{wang2021codet5} for the sake of training computation costs.
We utilized approximately 2.07 million paired PL-NL instances, encompassing six PLs: Java, Python, PHP, Javascript, Go, and Ruby. The total OpenAI API costs approximately \$2,000, and the rebuilt CodeSearchNet cgpt-CSN statistics are displayed in Table~\ref{tab:csn-stat}. 

In this study, we choose CodeT5~\cite{wang2021codet5}, a widely-used pre-trained model for source code based on an encoder-decoder framework similar to T5~\cite{raffel2020exploring}, as our pre-training model for downstream code intelligence tasks.
We follow Feng et al.~\cite{feng2020codebert} to employ cgpt-CSN to pre-train CodeT5, consisting of six PLs with unimodal and bimodal data. 
It is trained with four pre-training tasks, including a Masked Span Prediction (MSP) task, Identifier Tagging (IT), Masked Identifier Prediction (MIP), and Bimodal Dual Generation (BDG).  

As the pre-training implementation is not available, we re-implement the CodeT5 pre-training process based on Huggingface's T5~\cite{raffel2020exploring} PyTorch implementation, and the size of the model is 220M, same as CodeT5-base. We set the maximum source and target sequence lengths to be 512. We use the mixed precision of FP16 to accelerate the pre-training. We set the batch size to 48 and employed the peak learning rate 2e-5 with linear decay. Following the settings in CodeT5~\cite{wang2021codet5}, we pre-train the model with the denoising objective for 100 epochs and bimodal dual training for a further 50 epochs. 
In the fine-tuning phase, we follow their default settings for the hyperparameters in the CodeXGLUE~\cite{lu2021codexglue} tasks, such as learning rate, training steps, and batch size.

\subsubsection{Code Intelligence Tasks and Metrics}

For the code intelligence tasks, we cover four generation and one understanding tasks in the CodeXGLUE benchmark~\cite{lu2021codexglue} and employ the provided public datasets and the same data splits following it for all these tasks. We first consider two cross-modal, text-to-code and code-to-text generation tasks, two code-to-code generation tasks, and one code understanding task. 

\textbf{Code summarization} aims to summarize a function-level code snippet into natural language descriptions. The dataset comprises six PLs, including Ruby, JavaScript, Go, Python, Java, and PHP from CodeSearchNet~\cite{husain2019codesearchnet}. 
Empirical findings in Section~\ref{sec:evaluation} show that the USE metric aligns well with human evaluation. To provide a rich perspective view, we employ one reference-based metric USE~\cite{haque2022semantic} and one of our newly proposed reference-free metrics, MRR, to evaluate code summarization. Note that the reference-based metric USE refers to the comment generated by GPT-3.5-Turbo, as it provides better consistency with code.

\textbf{Code generation} is an NL-PL task that generates a code snippet based on NL descriptions. We employ the Concode dataset~\cite{iyer2018mapping} in Java, where the input contains both NL texts and class contexts, and the output is a Java function. We evaluate it with CodeBLEU~\cite{ren2020codebleu}, BLEU-4, and exact match (EM) accuracy that considers syntactic and semantic matches based on the code structure in addition to the n-gram match.

\begin{figure}[htbp]
\centerline{\includegraphics[scale=0.35]{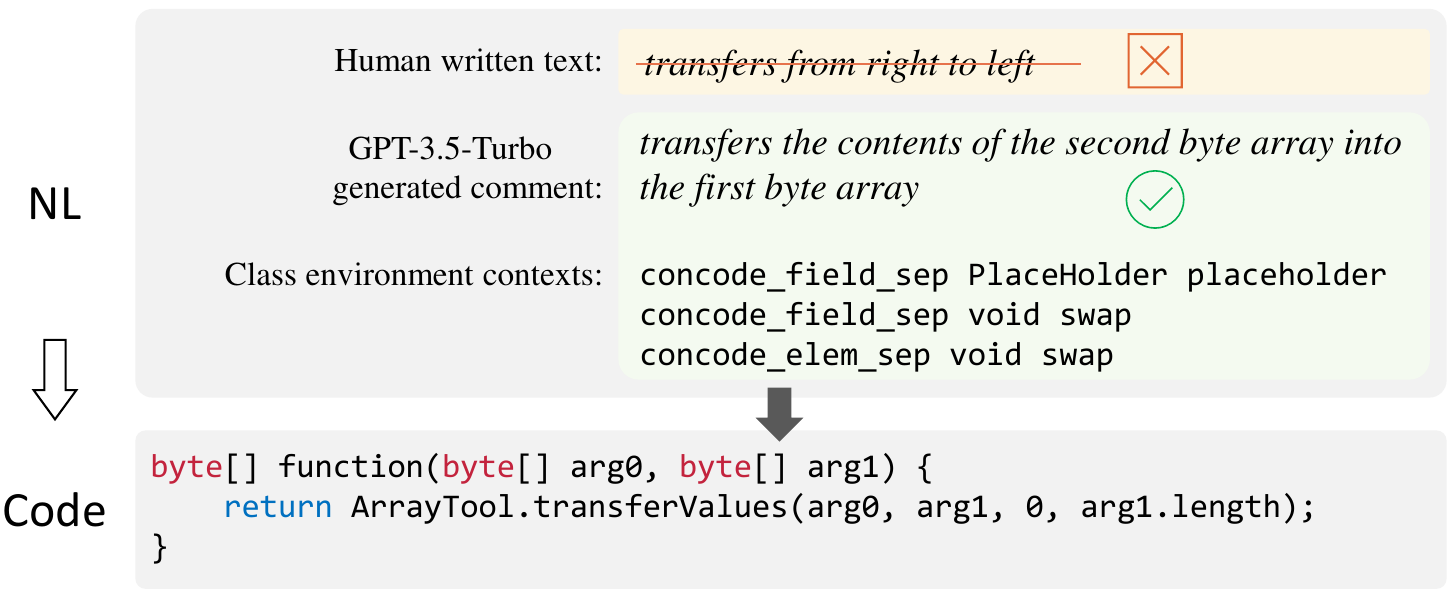}}
\caption{A Concode data sample rebuilt by GPT-3.5-Turbo.}
\label{fig-concode-rebuild}
\end{figure}

\textbf{Code translation} aims to migrate legacy software from one PL to another, where CodeXGLUE focuses on translating functions from Java to CSharp and vice versa. Because we omit C/CSharp in our updated version of the pre-training dataset, cgpt-CSN, we use the AVATAR~\cite{ahmad2021avatar}, a parallel corpus for Java-Python program translation, for code translation tasks.

\begin{table*}[]
\centering
\caption{Evaluation results of Code Summarization.}
\label{tab:eval-summarization}
\resizebox{\textwidth}{!}{
\begin{tabular}{c|c|cccccccccccc|cc}
\toprule
\multirow{2}{*}{pre-train} & \multirow{2}{*}{fine-tune} & \multicolumn{2}{c}{Javascript} & \multicolumn{2}{c}{PHP} & \multicolumn{2}{c}{Ruby} & \multicolumn{2}{c}{Java} & \multicolumn{2}{c}{Python} & \multicolumn{2}{c|}{Go} & \multicolumn{2}{c}{\textbf{Average}} \\
                           &                            & USE            & MRR           & USE        & MRR        & USE         & MRR        & USE         & MRR        & USE          & MRR         & USE        & MRR & USE        & MRR   \\ \hline
CSN                        & CSN                        & 0.5404         & 0.8102        & 0.6381     & 0.8102     & 0.5262      & 0.8189     & 0.6031      & 0.8083     & 0.5875       & 0.8294      & 0.6692     & 0.8580 & 0.5941 & 0.8225  \\
cgpt-CSN                   & CSN                        & 0.5420         & 0.8889        & 0.6429     & 0.8992     & 0.5375      & 0.8563     & 0.6111      & 0.9135     & 0.5911       & 0.9074      & 0.6763     & 0.9621 & 0.6002 & 0.9046   \\ 
CSN                        & cgpt-CSN-Sum                   & 0.7698         & 0.9623        & 0.8142     & 0.9202     & 0.7599      & 0.9477     & 0.8127      & 0.9342     & 0.6953       & 0.9529      & 0.8142     & 0.9120   & 0.7777 & 0.9382 \\
cgpt-CSN                   & cgpt-CSN-Sum                   & 0.7731         & 0.9665        & 0.8162     & 0.9617     & 0.7738      & 0.9617     & 0.8154      & 0.9553     & 0.6983       & 0.9552      & 0.8162     & 0.9244 & 0.7822 & 0.9541  \\ \hline
\multicolumn{2}{c|}{Human References}            & -              & 0.7710        & -          & 0.8094     & -           & 0.7988     & -           & 0.8091     & -            & 0.8153      & -          & 0.8481   & - & 0.8086\\
\multicolumn{2}{c|}{GPT-3.5-Turbo}           & -              & 0.9580        & -          & 0.9559     & -           & 0.9694     & -           & 0.9624     & -            & 0.9089      & -          & 0.9351  & - & 0.9483 \\ 
\bottomrule
\end{tabular}
}
\end{table*}

\textbf{Code refinement} is to detect which parts of code are buggy and fix them by generating a correct code sequence. We employ two Java datasets provided by Tufano et al.~\cite{tufano2019empirical} with various function lengths: small (fewer than 50 tokens) and medium (50-100 tokens). Due to the limited edit refining, there is a large token overlap between the source and target code. The EM measurement could better reflect the correctness of the refinement generation. We report EM and BLEU-4.  

\textbf{Clone detection} aims to measure the similarity between two code snippets and predict whether they have the same functionality. We experiment with the Java data provided by Wang et al.~\cite{wang2020detecting}. We employ Precision, Recall, and F1 scores for evaluating code clone detection. 


\subsection{Results and Analysis}
In this section, we evaluate the re-pre-trained CodeT5 on downstream tasks. For better illustration, cgpt-CSN denotes the model was pre-trained with an updated version of CodeSearchNet that we rebuilt using GPT-3.5-Turbo.

\textbf{Code Summarization} 
The initial two rows in Table~\ref{tab:eval-summarization} reveal that the model pre-trained with cgpt-CSN outperforms the one trained with vanilla CSN on code summarization task, as indicated by higher USE and MRR scores across all six PLs.
We further explore the effect of GPT-3.5-Turbo-generated comments data in the fine-tuning phase; we rebuild the CSN code summarization datasets, denoted as cgpt-CSN-Sum. Table~\ref{tab:eval-summarization} shows the results in the middle two rows. It should be noted that the ground truth for reference-based metrics USE on the rebuilt CSN summarization test set is GPT-3.5-Turbo-generated comments. We observe that the model fine-tuned with the rebuilt summarization data outperforms the model fine-tuned with human-referenced data.
These findings indicate that incorporating GPT-3.5-Turbo-generated data, both in the pre-training and fine-tuning phases, advances code summarization performance.

The final two rows in Table~\ref{tab:eval-summarization} compare MRR scores between reference comments and GPT-3.5-Turbo-generated comments on the CSN summarization test set. GPT-3.5-Turbo-generated comments achieve markedly higher MRR scores across all six PLs, aligning with findings from Section~\ref{sec:evaluation}. This consistency underscores the generalization capabilities of MRR metric across multiple programming languages.

\begin{table}[]
\centering
\caption{Evaluation results of NL-Code generation.}
\label{tab:eval-concode}
\resizebox{0.9\columnwidth}{!}{
\begin{tabular}{cc|ccc}
\toprule
pre-train & fine-tune    & CodeBLEU & BLEU & EM     \\ \hline 
CSN       & Concode      & 39.45 & 38.52 & 22.00     \\
CSN       & cgpt-Concode & 49.25 & 48.31 & 29.60     \\
cgpt-CSN  & Concode      & 40.56 & 40.91 & 22.10     \\
cgpt-CSN  & cgpt-Concode & 50.49 & 50.20 & 30.00     \\

\bottomrule
\end{tabular}
}
\end{table}
\textbf{Code Generation}
The vanilla input in the Concode dataset contains both NL text and class environment context. 
To investigate the impact of GPT-3.5-Turbo-generated comments in the fine-tuning stage, we also rebuild the Concode dataset by replacing the NL texts with GPT-3.5-Turbo-generated comments to form the updated inputs, as shown in Figure~\ref{fig-concode-rebuild} and the updated version noted as cgpt-Concode. 
Table~\ref{tab:eval-concode} shows that the model pre-trained with cgpt-CSN and fine-tuned with cgpt-Concode outperforms other training-tuning settings in three metrics. Compared to the model pre-trained on CSN and fine-tuned on Concode, GPT-3.5-Turbo-generated comments data in both pre-training and fine-tuning phases contributes to 11.04\%, 11.68\% and 8.00\% points improvement on CodeBLEU, BLEU, and EM respectively. Notably, in the fine-tuning phase, the ChatGPT enhanced cgpt-Concode contributes more significant performance gains than pre-training. 

We attribute this phenomenon to the characteristics of the NL-Code generation task itself. High-quality NL can better align the semantics with Code, and the model learns better representations in the embedding space during training. 
In the NL-Code generation task, the quality of NL directly determines the semantic correlation between the input and the final target code. Therefore, in the fine-tuning stage, improving the NL quality of the fine-tuning data of the NL-Code generation task can significantly improve the quality of the generated code. 
These results demonstrate that high-quality comment data generated from GPT-3.5-Turbo could significantly advance the performance of the NL-code generation task.

\begin{table}[]
\centering
\caption{Evaluation results of Code Translation.}
\label{tab:eval-translation}
\resizebox{0.9\columnwidth}{!}{
\begin{tabular}{cc|ccc}
\toprule
pre-train & fine-tune                    & CodeBLEU & BLEU  & EM   \\ \hline
CSN       & \multirow{2}{*}{Java2Python} & 48.28     & 51.29 & 2.57 \\
cgpt-CSN  &                              & 52.38     & 56.72 & 2.60 \\ \hline
CSN       & \multirow{2}{*}{Python2Java} & 55.52     & 56.82 & 1.21 \\
cgpt-CSN  &                              & 57.19     & 59.92 & 1.84 \\ 
\bottomrule
\end{tabular}
}
\end{table}

\textbf{Code Translation} 
The experiment results of the code translation task in the AVATAR test set are displayed in Table~\ref{tab:eval-translation}. The model pre-trained with cgpt-CSN data performs better than its counterpart pre-trained with human-commented CSN. Compared to the original CSN, GPT-3.5-Turbo-generated comments in the pre-training dataset contribute to an improvement of 4.10 CodeBLEU, 5.43 BLEU and 0.03 EM scores in Java-to-Python, 1.67 CodeBLEU, 3.10 BLEU and 0.63 EM scores in Python-to-Java translation, respectively. 
The code translation is an NL-unrelated task, and GPT-3.5-Turbo-generated NL in bimodal data can still improve performance. We attribute the improvement to the better alignment between PL and NL. The comments generated by GPT-3.5-Turbo preserve better semantic consistency with code, which could better representation for generation. 

\begin{table}[]
\centering
\caption{Evaluation results of Code Refinement.}
\label{tab:eval-refinement}
\resizebox{0.8\columnwidth}{!}{
\begin{tabular}{cc|cc}
\toprule
pre-train & fine-tune                       & EM    & BLEU  \\ \hline 
CSN       & \multirow{2}{*}{Refine-small}  & 21.41 & 77.41 \\
cgpt-CSN  &                                & 21.24 & 77.36 \\ \hline
CSN       & \multirow{2}{*}{Refine-medium} & 13.90 & 89.39 \\
cgpt-CSN  &                                & 13.74 & 89.51 \\ 

\bottomrule
\end{tabular}
}
\end{table}

\textbf{Code Refinement } Experiment results of EM and BLEU scores for code refinement tasks are shown in Table~\ref{tab:eval-refinement}. The results indicate no significant performance differences between models trained with CSN and cgpt-CSN on both small and medium code refinement test sets. These results imply that a deep comprehension of programming logic and syntax is essential for recognizing and correcting incorrect code patterns. The enhancement of NL comments in the pre-training stage is less directly applicable to identifying and correcting code errors. 

\textbf{Clone Detection} Table~\ref{tab:eval-clone-detection} 
presents F1 scores, Precision, and Recall for the clone detection task. Models pre-trained on CSN and cgpt-CSN show comparable performance, with F1-score differences of less than 0.5 percentage points. This none non-significant difference can be attributed to the fundamental nature of clone detection: the task primarily relies on analyzing syntactic and semantic similarities between code content pairs rather than their natural language descriptions. Consequently, while GPT-3.5-Turbo generates higher-quality comments, this enhancement does not transfer to improved clone detection performance, as the task is inherently more dependent on code-specific features than natural language documentation.

According to the experimental results on the five code intelligence tasks, we thus conclude that LLMs are qualified benchmark builders and answer the third research question(\textbf{RQ3}). 

\begin{table}[]
\centering
\caption{Evaluation results of Clone Detection.}
\label{tab:eval-clone-detection}
\begin{tabular}{cccc}
\toprule
pre-train  & F1     & P      & R     \\ \hline
CSN      & 0.9464  & 0.9455 & 0.9473 \\
cgpt-CSN & 0.9432  & 0.9358 & 0.9508 \\ 
\bottomrule
\end{tabular}
\vspace{0.01cm}
\end{table}
\newtcolorbox{RQ3-box}{
colframe = gray!50!black,
top = 2pt,                  
bottom = 2pt,               
left = 3pt,                 
right = 3pt,               
boxsep = 2pt}
\begin{RQ3-box}
\textbf{Ans. to RQ3: impact of rebuilt data for code tasks} 

Training data rebuilt using GPT-3.5-Turbo's high-quality comments significantly improves the performance of code intelligence tasks that rely on NL, like code summarization and NL-code generation, and it also enhances code translation capabilities. However, tasks primarily focused on code structure and semantics (code refinement and clone detection) show no meaningful improvement from enhanced comment quality.
\end{RQ3-box}
\section{Threats to validity}
\label{sec:threats}

\textbf{Internal validity}. 
The scope of our study's conclusions is constrained by limitations in computing resources and the expenses related to using the OpenAI API. Consequently, our research focused solely on a specific pre-trained code model, CodeT5, and we did not extend our analysis to include open-source Code LLMs such as Code Llama~\cite{roziere2023code}.

Our study focuses on demonstrating the superior quality of LLM-generated comments over human references for improving training datasets, rather than achieving SOTA performance in code intelligence tasks. While current LLMs might outperform traditional pre-trained models on these tasks, our work's primary contribution lies in dataset quality enhancement rather than model performance comparison. We acknowledge that a direct performance comparison between our rebuilt pre-trained models and current LLMs could provide additional insights, but this falls outside our core research objective of improving training data quality for future model development.

\textbf{External validity}. 
LLMs' outputs can vary significantly with slight changes in the prompt structure, wording, or context. Our study employs a simple, consistent prompt across programming languages to ensure reproducibility. However, this approach may not capture the full potential of LLMs, which could be achieved through more sophisticated prompt engineering. Another threat to the validity stems from the versions of ChatGPT used in our research, specifically Text-davinci-003, GPT-3.5-turbo, and GPT-4-Turbo, which represent ChatGPT's capabilities at a certain point in time. As commercial LLMs undergo continuous updates, the performance characteristics and findings reported in this study may evolve with newer versions.

\section{Related Works}

Recent advancements in code intelligence tasks are heavily dependent on the quality of code-comment paired data. Previous studies have identified significant issues in code documentation~\cite{linares2015developers, wen2019large, ibrahim2012relationship, malik2008understanding}. Fluri et al. found that 3-10\% of code comment changes lagged behind the corresponding code changes in seven Java open-source projects ~\cite{fluri2007code}. 
Such obsolete comments may provide misleading information to developers, leading them to write vulnerable code~\cite{ibrahim2012relationship} and thus degrading the quality of the software. The noisy code comments data could degrade the performance of data-driven-based learning models for code intelligence tasks~\cite{sun2022importance}.

The main focus of the research community is on developing customized models that can unleash the value of the available data in specific tasks. As mentioned in~\cite{zhao2021impact, liu2020simplifying}, improving the quality of the training data is still a research opportunity for machine learning, including DL-based source code models. Sun et al.~\cite{sun2022importance} proposed the first framework to improve the dataset quality for code search datasets. Their data cleaning framework, which consists of two subsequent filters, a rule-based syntactic filter and a model-based semantic filter, is considered to filter the noisy data. Xu et al.~\cite{xu2023data} investigated the data quality issue in the obsolete comment detection problem by proposing data cleaning and adversarial learning techniques. They found that the performance of DL models does improve with the cleaned training data. Unlike the above studies, our work tackles the data quality issue by rebuilding the dataset via LLMs, thereby replacing the noisy data with high-quality ones. 

\section{Conclusion}
We present a comprehensive evaluation of LLM-generated code comments against human-written references using novel reference-free metrics based on inconsistency detection and code search tasks. Our empirical study found that LLM-generated comments are superior to comments written by humans in the original repositories, challenging the conventional use of human references as the gold standard for code summarization. This finding led us to reconstruct the CodeSearchNet dataset using LLM-generated comments and subsequently retrain the CodeT5 model. The rebuilt model showed significant improvements across multiple code intelligence tasks, particularly in natural language-related tasks such as code summarization and NL to Code generation. These results validate both the superior quality of LLM-generated comments and their practical value in enhancing training datasets for code intelligence models.


\bibliographystyle{ieeetr}
\bibliography{main_references}

\end{document}